\documentclass[aps]{revtex4}

\usepackage[utf8]{inputenc}
\usepackage[T1]{fontenc}

\usepackage{amsmath}
\usepackage{amssymb}
\usepackage{amsfonts}
\usepackage{color}
\usepackage{mathtools}
\usepackage[scr=boondox]{mathalfa}

\newcommand{\period}{\hspace{0.005\linewidth} . \hspace{0.02\linewidth} }
\newcommand{\coma}{\hspace{0.02\linewidth} , \hspace{0.02\linewidth} }
\newcommand{\with}{\hspace{0.02\linewidth} \text{ with } \hspace{0.02\linewidth} }
\newcommand{\andd}{\hspace{0.02\linewidth} \text{ and } \hspace{0.02\linewidth} }
\newcommand{\for}{\hspace{0.02\linewidth} \text{ for } \hspace{0.005\linewidth} }

\newcommand{\etal}{{\em et al.~}}

\newcommand{\rey}[1]{\overline{#1}}
\newcommand{\mean}[1]{\langle{#1}\rangle}
\newcommand{\bs}[1]{\boldsymbol{#1}}
\newcommand{\x}{\bs{x}}
\newcommand{\R}{\bs{r}}
\newcommand{\ud}{\mathrm{d}}
\newcommand{\prll}{\parallel}

\newcommand{\vi}{v^{d}}
\newcommand{\vs}{v^{s}}
\newcommand{\F}{\Pi}
\newcommand{\pis}{\pi^{s}}
\newcommand{\Ret}{\mathtt{Re}_\mathtt{t}}
\newcommand{\halfp}[1]{\{#1\}_{\frac{1}{2}}}
\newcommand{\msh}[1]{\left.\rey{#1}\right|_\mathtt{sh}}
\newcommand{\mns}[1]{\left.\rey{#1}\right|_\mathtt{ns}}
\newcommand{\Ps}{P_\mathtt{sh}}
\newcommand{\Einj}{\mathscr{E}_\mathrm{inj}}
\newcommand{\flt}[1]{{\left\langle{#1}\right\rangle_\ell}}
\newcommand{\fflt}[1]{{\big.\widetilde{#1}\big|_\ell}}
\newcommand{\res}[1]{{\left.{#1}\right|_\ell^\mathrm{res}}}
\newcommand{\sub}[1]{{\left.{#1}\right|_\ell^\mathrm{sub}}}
\newcommand{\resz}[1]{{\left.{#1}\right|_0^\mathrm{res}}}
\newcommand{\subz}[1]{{\left.{#1}\right|_0^\mathrm{sub}}}
\newcommand{\ci}{\mathcal{C}}
\newcommand{\jj}{j}
\newcommand{\jjv}{\bs{\jj}}
\newcommand{\pc}{\mathcal{P}_c}

\newcommand{\rh}{\rey{\rho}}
\newcommand{\eps}{\varepsilon}
\newcommand{\epsr}{\rey{\varepsilon}}

\date{\today}

\begin{document}

\title{
Cascade of circulicity in compressible turbulence
}

\author{O. Soulard} \email{olivier.soulard@cea.fr}
\author{A. Briard}
\affiliation{CEA, DAM, DIF, F-91297 Arpajon, France}

\begin{abstract}
The purpose of this work is to investigate whether a cascading process can be associated with the rotational motions of compressible three-dimensional turbulence.
This question is examined through the lens of circulicity, a concept related to the angular momentum carried by large turbulent scales. 
By deriving a Monin-Yaglom relation for circulicity, we show that an ``effective'' cascade of this quantity exists, provided the flow is stirred with a force having a solenoidal component.
This outcome is obtained independently from the expression of the equation of state. 
To supplement these results, a coarse-graining analysis of the flow is performed. This approach allows to separate the contributions of the transfer and production terms of circulicity and to discuss their respective effects in the inertial range.
\end{abstract}

\maketitle

\vspace{-4ex}
%=======================================================================
\section{Introduction}
%=======================================================================
In incompressible turbulence,  the behavior of small inertial scales is commonly described by invoking the physical image of the Richardson cascade \cite{richardson1922,davidson2011,alexakis2018}: 
under the stirring action of velocity gradients,  inertial eddies break down into smaller and less energetic ones, that  break down into even smaller and less energetic ones, and so on until viscous dissipation occurs.
This phenomenological description is intimately linked with two sets of formal results stemming respectively from Kolmogorov's \cite{kolmogorov1941,frisch1995} and Onsager's \cite{onsager1949, eyink2006} works.

The first approach focuses on the statistical properties of the velocity field taken at different points, a technique sometimes called ``point-splitting''. 
Kolmogorov showed that, in the inertial range of homogeneous turbulence,  kinetic energy flows on average from large to small scales with a constant flux given by its mean dissipation rate $\epsr$.
When turbulence is isotropic, this result takes the form of Kolmogorov's  $4/5^\mathrm{th}$ law, while in anisotropic turbulence, it is expressed with the Monin-Yaglom relation \cite{monin1975}. 
Both expressions relate the triple correlation of the two-point increments of the velocity field to $\epsr$.
They are both derived from the incompressible Navier-Stokes equations under a crucial assumption : the mean dissipation  $\epsr$  must remain finite when the Reynolds number  $\Ret$ tends to infinity.
This hypothesis is sometimes referred to as the zero$^\text{th}$ law of turbulence and is related to the notion of dissipative anomalies \cite{taylor1935,kolmogorov1941,kolmogorov1941b,onsager1949,john2021}.

Dissipative anomalies are also at the heart of the second set of formal results, the one originating from Onsager's work.
The concept of anomaly arises from the observation that, when the Reynolds number $\Ret$ increases,  dissipative events take place in thinner and thinner structures while becoming more and more intense, in such a way that their average $\epsr$ is finite. 
In the limit $\Ret \to \infty$, several authors, starting with Onsager \cite{onsager1949,eyink2006,saw2016,dubrulle2019}, conjectured that this phenomenology could be described in the form of a weak singular (``anomalous'') solution of  Euler equations.
As opposed to Kolmogorov's approach, Onsager's does not necessarily require statistical averages of two-point quantities. Instead, it can be expressed with the local properties of a regularized ``coarse-grained'' velocity field.
Despite their inherent differences, Onsager's coarse-graining and Kolmogorov's point-splitting visions share the same core phenomelogical ideas. Accordingly, the coarse-graining approach has given rise to a series of predictions closely related to Kolmogorov's  $4/5^\mathrm{th}$ law and to the idea that energy is transferred from scale to scale at a constant rate \cite{eyink2006}. 
Both the ``point-splitting'' and ``coarse-graining'' approaches sustain the idea that an energy cascade takes place in the inertial range of incompressible turbulence.

When density inhomogeneities or compressible effects are present, the physical image of the Richardson cascade is harder to uphold.
In that case, eddies are not only stirred and broken by velocity gradients, they are also affected by baroclinic production and by local dilatations and compressions. 
In principle, these additional processes may very well lead to an inverse transfer of energy, at the very least locally and transiently.
These processes also cast their shadow on the point-splitting and coarse-graining approaches.
 Their adaptations to compressible turbulence face several hurdles, some of which have  only been recently overcome. 
In particular,  Aluie \cite{aluie2011,aluie2012,aluie2013} and Eyink \& Drivas \cite{eyink2018} have produced two seminal works based on the coarse-graining approach. They have  shown that in compressible turbulence, an inertial range exists where a local conservative cascade of kinetic energy takes place,  despite the fact that kinetic energy is no longer an inviscid invariant of the flow. This result is conditioned on the presence of dissipative anomalies and on the sufficient roughness of the turbulent fields. 

From the point-splitting side, no equivalent predictions have so far been derived. Still, several adaptations of the Monin-Yaglom relation to compressible turbulence have been proposed \cite{falkovich2010,galtier2011,wagner2012,banerjee2013,kritsuk2013,banerjee2014,banerjee2017,banerjee2018,andres2018,andres2019,ferrand2020,simon2021}.
But as opposed to the coarse-graining approach, most these efforts were not aimed at proving the existence of a cascading process. Instead, they were mostly driven by a practical goal: being able to evaluate the dissipation and large-scale injection of energy knowing two-point correlations.
This information is indeed useful for interpreting several astrophysical phenomena \cite{andres2019,simon2021}.
As it turns out, one of the main difficulties for establishing Monin-Yaglom relations in compressible flows comes from the pressure field and the way it is coupled with other thermodynamical variables.
To alleviate this difficulty, most of the mentioned works \cite{falkovich2010,galtier2011,wagner2012,banerjee2013,kritsuk2013,banerjee2014,banerjee2017,banerjee2018,andres2018,andres2019,ferrand2020,simon2021} make a simplifying assumption on the thermodynamical state of the flow: it is either isothermal \cite{galtier2011,wagner2012,banerjee2013,kritsuk2013,banerjee2017,banerjee2018,andres2018,andres2019,ferrand2020}, polytropic \cite{falkovich2010,banerjee2014} or isentropic \cite{simon2021}. These assumptions are not innocuous: they de facto prevent the existence of several physical phenomena. For instance, the first two assumptions suppress the baroclinic torque and the development of convective instabilities such as the Rayleigh-Taylor and Richtmyer-Meshkov ones \cite{zhou2017a,zhou2017b,soulard2012b,soulard2017}. Besides, the third one is not compatible with the entropy cascade predicted in \cite{eyink2018}.
Therefore, despite their great practical use, the Monin-Yaglom relations derived so far are in principle restricted to particular flows.
In this regard, it is worth highlighting  that all the mentioned Monin-Yaglom relations are coupled to the energetic and mixing content of the flow, either through the presence of the pressure field, either because they purposely consider the inertial transfer of the total energy. None of them is purely dynamical, in the sense of involving only velocity and density correlations.
Whether such a relation exists remains an open question. But its answer is tightly linked with the possibility of finding a Monin-Yaglom equation that is not restricted to a particular thermodynamical assumption. Besides, it can already be stressed that, if it exists, this answer would require finding a cascading quantity different from the energy: as shown with the coarse-graining approach in \cite{aluie2011,aluie2012,aluie2013,eyink2018}, the pressure is indeed one of the components acting in the interscale transfer of kinetic energy.

Before exploring further these issues, let us briefly return to the incompressible case. In this context, the velocity field is purely solenoidal and is directly connected to the vorticity field through a Helmholtz decomposition \cite{sagaut2018}.
In homogeneous turbulence, this connection is such that the kinetic energy spectrum becomes equal to the  circulicity spectrum.
Circulicity is a term coined by Kassinos \etal \cite{kassinos2001} in order to refer to a tensor describing the ``large-scale'' structure of the vorticity field. 
In this work, we will slightly tweak this definition and use circulicity to refer to the ``large-scale'' structure of the angular momentum.
Indeed,  when density is constant, vorticity becomes a proxy for measuring the angular momentum of fluid particles, so that the two notions become indistinguishable.
Piecing these elements together, one arrives at the conclusion that the energy cascade predicted by the point-splitting and coarse-graining approaches also coincides with a cascade of circulicity.
This co-occurence agrees with the physical picture of the Richardson cascade in which energy is transferred to smaller scales by the motion of turbulent eddies, which can be  thought of as coherent regions of angular momentum.

Coming back to compressible flows, the velocity field now possesses a dilatational component and is not solely representative of vortical motions. But most importantly, vorticity itself loses one of its key physical meaning. 
It is not connected to the angular momentum of fluid particles any longer.
Indeed, angular momentum is defined with respect to the curl of the linear momentum and is only related to vorticity when density is constant \cite{soulard2020,soulard2022}. 
Therefore, contrary to the incompressible case, the spectral properties of  kinetic energy and circulicity  dissociate in compressible turbulence. 
In particular, in homogeneous turbulence, the identity between the circulicity and kinetic energy spectra ceases to be valid. Instead, the circulicity spectrum becomes equal to the spectrum of the solenoidal component of the linear momentum. The latter is indeed the Helmholtz dual of the angular momentum, just as the solenoidal velocity is with respect to vorticity.
As a consequence of this dissociation, the existence of the energy cascade predicted with the coarse-graining approach in \cite{aluie2011,eyink2018} does not automatically extend to circulicity : even if energy cascades to small scales, it is not known whether this transfer goes along with the apparition of smaller whirls.
Thus, the question arises as to whether a circulicity cascade exists  in compressible turbulence. 
This question sends us back to the issues we already raised when commenting existing Monin-Yaglom relations. Indeed, as shown in \cite{soulard2020,soulard2022}, the evolution of the angular momentum and of the solenoidal linear momentum are not directly impacted by the pressure field.
Hence, the study of their properties opens a door for deriving a Monin-Yaglom relation that is not bound by the equation of state and that is decoupled from the thermodynamics and mixing content of the flow.

Given these considerations, the purpose of this work is to study the transfer of circulicity in the inertial range of a compressible flow. More particularly, we aim to discuss whether a circulicity cascade exists and, if it does, what it implies for inertial scales.
This study is amenable to both the point-splitting and coarse-graining approaches. For the sake of completeness, we will use the two techniques. 
To begin with, we will derive a Monin-Yaglom relation for the solenoidal component of the linear momentum, an appellation hereafter abridged to solenoidal momentum. We will then discuss how this relation depends on the different variables describing the flow. We will also discuss how it may potentially provide information about the scaling of the turbulent spectra.
These different elements will also be examined through the lens of the coarse-graining approach. To this end, spatial filtering will be applied to the solenoidal momentum equation.

%=======================================================================
\section{Governing Equations}
%=======================================================================

%-----------------------------------------------------------------------
\subsection{Compressible Navier-Stokes equations}
%-----------------------------------------------------------------------
We consider a flow which state is defined by its density $\rho$, velocity $\bs{v}$, internal energy $e$ and species mass fractions $\left\{ c_\alpha, \; \alpha=1\cdots N_s \right\}$. These variables evolve according to the Navier-Stokes equations \cite{giovangigli1999}:
\begin{subequations}\label{eq:NS_cons}
\begin{align}
\label{eq:ns_rho}
 \partial_t \rho + \partial_j(\rho v_j) &= 0
\coma
\\
\label{eq:ns_rhou}
 \partial_t (\rho v_i) + \partial_j(\rho v_i v_j) &= - \partial_j ( p \delta_{ij} + \sigma_{ij} ) + f_i
\coma
  \\
\label{eq:ns_rhoe}
 \partial_t (\rho e) + \partial_j(\rho v_j e) &= \rho \eps - p \partial_j v_j - \partial_j\sigma^{(e)}_j
\coma
\\
\partial_t (\rho c_\alpha ) + \partial_j(\rho v_j c_\alpha) & = - \partial_j\sigma^{(\alpha)}_j
\end{align}
where  $p$ is  the pressure, $\bs{f}$ a volumetric force, $\bs{\sigma}$  the viscosity tensor, $\bs{\sigma}^{(e)}$  the molecular heat flux,  $\bs{\sigma}^{(\alpha)}$ the molecular diffusion flux of species~$\alpha$ and $\rho \eps= -\sigma_{ij} \partial_jv_i$  the dissipation rate of the kinetic energy.
For simplicity, we assume that the viscosity tensor is defined by:
\begin{align}
\sigma_{ij} = - \mu S_{ij} - \eta \partial_k v_k \delta_{ij}
\with 
S_{ij} =\partial_jv_i + \partial_iv_j - \frac{2}{3} \partial_k v_k \delta_{ij}
\coma
\end{align}
where the shear viscosity $\mu$ and the bulk viscosity $\eta$ are constant.
The molecular heat and diffusion fluxes, $\bs{\sigma}^{(e)}$ and $\bs{\sigma}^{(\alpha)}$, will not be used in the remainder of this text and do not need to be detailed. Expressions for these quantities can be found for instance in \cite{giovangigli1999}.
To close the flow description, we assume that the pressure is given by an equation of state which is a function of density, energy and mass fractions:
\begin{align}  \label{eq:eos}
p \equiv p(\rho,e,\bs{c})
\period
\end{align}
This function can be general and is not constrained by an isothermal, polytropic or isentropic assumption. Similarly to $\bs{\sigma}^{(e)}$ and $\bs{\sigma}^{(\alpha)}$, the equation of state does not play any role in the forthcoming analysis and does not need to be made explicit.
\end{subequations}

%-----------------------------------------------------------------------
\subsection{Helmholtz decomposition of the momentum}
%-----------------------------------------------------------------------
Using an Helmholtz decomposition, along with the fact that the domain considered in this study is unbounded,  the momentum $\jjv = \rho \bs{v}$ can be split into a solenoidal and a dilatational component respectively denoted by $\bs{s}$~and~$\bs{d}$ \cite{soulard2020,soulard2022}:
\begin{align} \label{eq:q_decomp}
\jjv = \rho \bs{v} = \bs{s}+ \bs{d}
\with 
d_i = - \partial_i \Phi \andd
s_i = \epsilon_{ijk} \partial_j \Psi_k
\coma
\end{align}
where $\epsilon_{ijk}$ is the Levi-Civita tensor.
The scalar and vector potentials $\Phi$ and $\bs{\Psi}$ are given by:
\begin{align} \label{eq:q_pot}
\partial^2_{jj} \Phi = -\Xi 
\andd
\partial^2_{jj} \Psi_i = - \Omega_i
\coma
\end{align}
where $\Xi$ is divergence of the momentum and $\bs{\Omega}$ is the angular momentum of a fluid particle:
\begin{align}
\Xi = \partial_j(\rho v_j) = -\partial_t \rho \andd
\Omega_i = \epsilon_{ijk} \partial_j (\rho v_k)
\period
\end{align}
It is important to stress that the angular momentum $\bs{\Omega}$ does not carry the same information as  the vorticity $\bs{\omega}$. The latter is defined by:
$$
\omega_i = \epsilon_{ijk} \partial_j v_k = {\Omega_i}/{\rho} - \epsilon_{ijk}v_k \partial_j \rho/\rho
\period
$$
Similarly,  the momentum divergence $\Xi$ is different from the velocity divergence $\Theta$ defined as
$$
\Theta = \partial_j u_j = {\Xi }/{\rho} - u_j \partial_j \rho/\rho
\period
$$
Accordingly, the Helmholtz decomposition of the momentum $\rho \bs{v}$ is also different from the Helmholtz decomposition of the velocity field $\bs{v}$, which  can be written as:
\begin{align}
  \bs{v} = \bs{\vs} + \bs{\vi} 
\with  
\vi_i = - \partial_i \varphi^v 
\coma
\vs_i = \epsilon_{ijk} \partial_j \psi^v_k
\andd 
\partial^2_{jj} \varphi^v = - \Theta
\coma
\partial^2_{jj} \psi^v_i = - \omega_i
\period
\end{align}
Except when density is constant, one generally has:
$$
\bs{\Omega} \ne \rho \bs{\omega} \coma 
\Xi \ne \rho \Theta \coma
\Phi \ne \rho \varphi^v \andd \bs{\Psi} \ne \rho \bs{\psi}^v
\period
$$

%-----------------------------------------------------------------------
\subsection{Circulicity} \label{sec:circ}
%-----------------------------------------------------------------------
Given the Helmholtz decomposition \eqref{eq:q_decomp}-\eqref{eq:q_pot}, the solenoidal momentum $\bs{s}$ characterizes the rotational motions of the flow, ``rotational'' being used here as a reference to angular momentum and not vorticity.
This link between $\bs{s}$ and $\bs{\Omega}$ can be formalized further with the notion of circulicity. More precisely, in this work, we define the local value of the circulicity field $\ci$ by:
\begin{align}
\ci = \frac{1}{2} s_is_i
\period
\end{align}
The quantity $\ci/\rho$ can be interpreted as the contribution of rotational motions to the kinetic energy $\frac{1}{2} \rho v_iv_i$. 
But $\ci$ also has another interpretation that is not linked to energy and that is more easily discussed in homogeneous turbulence.
In this context,  the mean of $\ci$, which is half the variance of $\bs{s}$, is equal to the integral of the solenoidal momentum spectrum $E_{ss}$:
\begin{align} \label{eq:ciess}
\text{Homogeneous turbulence : } \hspace{0.05\linewidth}
\rey{\ci} = \frac{1}{2} \rey{s_is_i} = \int E_{ss} \ud k
\coma
\end{align}
where $\rey{\cdot}$ denotes the ensemble mean and where $k$ is the wave-number.
Besides, given the Poisson equations \eqref{eq:q_pot}, the spectrum $E_{ss}$ of $\bs{s}$ is linked to the spectrum $E_{\Omega\Omega}$ of $\bs{\Omega}$ by :
\begin{align} \label{eq:ess}
\text{Homogeneous turbulence : } \hspace{0.05\linewidth}
E_{ss} = \frac{E_{\Omega\Omega}}{k^2}
\period
\end{align}
 Equation \eqref{eq:ess} indicates that the solenoidal momentum carries the same two-point information as the angular momentum,  with the difference that a greater weight is given to large scales. 
When combined with Eq. \eqref{eq:ciess}, this result shows that the mean circulicity $\rey{\ci}$ accounts for the large-scale content of the angular momentum. 
 This conclusion has been illustrated for homogeneous turbulence, but the overall correspondence between circulicity and large-scale angular momentum extends to the inhomogeneneous case.

Note also that in a constant-density flow, one would simply have  $\rey{\ci} = \rho ^2 \rey{v_iv_i}/2$.
Hence,  circulicity would be redundant with  kinetic energy  and there would be no need to study its properties, nor those of the solenoidal momentum two-point correlations.
However,  this is no longer the case when density is variable. As mentioned in the introduction, the concepts of circulicity and energy dissociate in compressible flows. 
Finally, it is worth stressing that the notion of circulicity used here is only loosely based on the one proposed by Kassinos \etal  \cite{kassinos2001}. The latter introduced this concept to measure different contributions of the  Reynolds-stress tensor of an incompressible velocity field.
Here, we only preserved the idea that circulicity gives access to the large-scale structure of the angular momentum.

%-----------------------------------------------------------------------
\subsection{Evolution of the solenoidal momentum $\bs{s}$}
%-----------------------------------------------------------------------
The solenoidal momentum $\bs{s}$ is the main focus of this work because of its connection to angular momentum and circulicity. To derive its evolution, we start from the linear momentum equation \eqref{eq:ns_rhou} and note that $\partial_t d_i = - \partial_i \partial_t \Phi$.  This term can then be added along with $\partial_ip$ into the gradient of a pseudo-pressure $\pis$, which value is set by the fact that $\bs{s}$ is divergence-free. The same can also be done for the dilatational parts of the forcing and viscous terms. This procedure yields:
\begin{subequations} \label{eq:sol_mom}
\begin{align}
\label{eq:ns_qs}
\partial_t s_i + \partial_j\left( \rho v_i v_j  \right) = - \partial_i \pis - \partial_j\sigma^s_{ij} + f_i^s
\coma
\end{align}
where  $\bs{\sigma}^s$ and $\bs{f}^s$ are the solenoidal parts of the forcing and viscous terms:
\begin{align}
\sigma_{ij}^s = - \mu \left( \partial_j v_i + \partial_i v_j\right) + 2 \mu \partial_k v_k \delta_{ij}
\andd
f_i^s = \epsilon_{ijk} \partial_j \psi_k^f \with \partial^2_{jj} \psi_i^f = - \epsilon_{ijk} \partial_j f_k
\period
\end{align}
The pseudo-pressure $\pis$  enforces the divergence-free constraint of $\bs{s}$  for the advection term $\partial_j(\rho v_i v_j)$:
\begin{align}\label{eq:poisson}
\partial^{2}_{jj} \pis = -\partial^2_{ij}\left( \rho v_i v_j \right)
\period
\end{align}
\end{subequations}
It is worth highlighting that $\pis$ is  different from the actual pressure $p$:
$$
\pis \ne p
\period
$$
The pressure $p$ is set by the equation of state \eqref{eq:eos}. By contrast, $\pis$ is the solution of a Poisson equation and is set by the dynamics of the flow. This remains true whether the flow displays strong compressibility effects or not.
Another significant point is that $p$ does not appear explicitly in Eq. \eqref{eq:ns_qs}. Thus, even though the equation of state is important for determining the overall evolution of the flow, its influence on the solenoidal momentum $\bs{s}$ is only indirect. This is not the case for all solenoidal fields. For instance, pressure appears explictly in the evolution equation of the solenoidal velocity field $\bs{v}^s$.

%-----------------------------------------------------------------------
\subsection{Evolution of the circulicity $\ci$}
%-----------------------------------------------------------------------
Starting from the evolution equation \eqref{eq:ns_qs} of the solenoidal momentum and using the Poisson equations \eqref{eq:q_pot}, the following evolution of the circulicity can be deduced:
\begin{align} \label{eq:circ}
\partial_t \ci + \partial_j \mathcal{F}_j^c =   \pc  - \eps_c  + s_if_i^s
\coma
\end{align}
where $\bs{\mathcal{F}}^c$ is the flux of circulicity:
\begin{align}
\mathcal{F}_i^c = \frac{1}{2} | \jjv|^2 v_i  + \pis \jj_i + \Phi \partial_j\left( \rho v_iv_j + \pis \delta_{ij}  \right) +    \mu s_j(\partial_jv_i - \partial_i v_j) 
\coma
\end{align}
where $\pc$ and $\eps_c$ are terms associated with the local production and dissipation of circulicity:
\begin{align}
\pc =  - \frac{1}{2} | \jjv|^2 \Theta + \pis \Xi
\andd
\eps_c = \mu \Omega_i\omega_i
\coma
\end{align}
 where we recall that $\jjv = \rho \bs{v}$, $\Theta = \partial_i v_i$ and $\Xi = \partial_i (\rho v_i)$.

This formulation of the circulicity equation allows to highlight the role played by the dilatational components of the velocity and momentum: it is the divergences of these fields that control  the circulicity production term $\pc$. 
The other interest of this formulation is to allow for a direct comparison with the evolution of the total momentum norm. Indeed, one has:
\begin{align} \label{eq:j2}
\partial_t \left(\frac{1}{2}| \jjv|^2 \right)  + \partial_j\left[\frac{1}{2} | \jjv|^2 v_j  + p \jj_j + \sigma_{ij}  \jj_i \right] =    - \frac{1}{2} | \jjv|^2 \Theta + p \Xi + \sigma_{ij} \partial_j \jj_i  + \jj_i f_i
\period
\end{align}
The right-hand sides of equations \eqref{eq:circ} and \eqref{eq:j2} share the same structure, with similar source, dissipation and forcing terms.
The main differences are that, in Eq. \eqref{eq:circ}, the dynamic pressure $\pis$ replaces the actual pressure $p$ and that the solenoidal force replaces the total one.
Besides,  the dissipation term $\eps_c$ only accounts for shear viscosity effects and not bulk ones.
It is worth stressing that $\eps_c$ is not necessarily positive. The angular momentum and vorticity fields can indeed be misaligned.
This misalignment is linked to the density and velocity gradients as follows:
$$
\eps_c = \mu \,\rho \omega_i \omega_i + \mu\, v_i \left(\partial_jv_i - \partial_i v_j \right) \partial_j \rho 
\period
$$
However, if we assume that the density and velocity fields decorrelate at small scales, the mean of $\eps_c$ will be dominated by the mean of the first term. In that case, the mean of $\eps_c$ remains positive, even though local values of $\eps_c$ can be negative. In the remaining of this work, we will assume that such an hypothesis holds and that:
$$
\rey{\eps_c} > 0
\period
$$

%=======================================================================
\section{Monin-Yaglom relation and transfer of circulicity  in homogeneous turbulence}
%=======================================================================

Our purpose in this section is to study how circulicity is transferred from scale to scale in the inertial range of a homogeneous compressible flow.
Given the relation between circulicity and solenoidal momentum (see Sec. \ref{sec:circ}), we start by deriving a Karman-Howarth equation for the solenoidal momentum $\bs{s}$. Then, we use this equation to express a Monin-Yaglom relation describing the interscale transfer of this quantity and by extension of circulicity.
This relation is compared against existing ones \cite{falkovich2010,galtier2011,wagner2012,banerjee2013,kritsuk2013,banerjee2014,banerjee2017,banerjee2018,andres2018,andres2019,ferrand2020,simon2021} and is also used to discuss inertial range scalings.
This section only makes use of the point-splitting approach. The coarse-graining method will be discussed in the next section.

%-----------------------------------------------------------------------
\subsection{Karman-Howarth equation for the solenoidal momentum}
%-----------------------------------------------------------------------

The first step in our study consists in deriving a Karman-Howarth equation for the solenoidal momentum, or, in other words,  an evolution equation for the second-order structure function of the solenoidal momentum.
The latter is defined by:
$$
\rey{\Delta s_i \Delta s_i} (\R,t) = 2 \rey{s_i s_i} - 2 \rey{s_i(\x)s_i(\x+\R)}
\coma
$$
where for any quantity $X$, $\Delta X$ refers to the  difference between the values of $X$ taken at two different points $\x$ and $\x'$, separated by the vector~$\bs{r}= \x' - \x$. For later purposes, it will also be useful  to introduce the median value $\halfp{X} $of $X$ at these two points. These two-point difference and median values are defined by:
$$
\Delta X(\x,\R,t) = X(\x',t) - X(\x,t) \andd \halfp{X}(\x,\R,t) =  (X(\x',t) + X(\x,t))/2
\with \R = \x'-\x
\period
$$
We would like to stress again that $\rey{\Delta s_i \Delta s_i}$ is related to circulicity and that, as such, it accounts for the large-scale content of the angular momentum. Indeed, one has:  $\rey{\Delta s_i \Delta s_i} (\R,t) = 4 \rey{\ci} - 4 \int e^{\imath \bs{k}\cdot \bs{r}} \frac{E_{\Omega \Omega}}{2 \pi k^4} \ud \bs{k}$. 
Therefore, studying the inertial range transfer of $\rey{\Delta s_i \Delta s_i}$ is equivalent to studying the transfer of circulicity.

Starting from Eq. \eqref{eq:ns_qs}  and using some of the relations detailed in App. \ref{sec:useful}, we find after some algebra that:
\begin{subequations}
\begin{align}
\label{eq:dsds}
& \partial_t \rey{\Delta s_i \Delta s_i}(\R,t) =
- \partial_{r_j}  \mathscr{F}_j(\R,t) - 4 \Einj(t)  + 2 \mu \partial^{2}_{r_jr_j} \rey{\Delta v_i \Delta s_i}(\R,t)
+ 2 \rey{\Delta s_i \Delta f_i^s}(\R,t)
\coma
\end{align}
where the circulicity flux $\bs{\mathscr{F}}$ and  injection rate $\Einj$ are defined by:
\begin{align}
 & \label{eq:flux}
\mathscr{F}_j(\R,t)=  
\rey{\Delta v_j \Delta \jj_i\Delta\jj_i} (\R,t) - 2  \rey{\Delta v_j^d \halfp{\jj_i\jj_i}} (\R,t)- 4 \rey{\Delta {d}_i  \halfp{\rho  v_i v_j}}(\R,t)
\coma
\\
\label{eq:EE}
\andd & \Einj(t) =   \rey{s_if_i^s} - \partial_t\rey{\ci} =
\rey{\eps_c}(t) - \rey{\pc}(t)
\period
\end{align}
\end{subequations}
The first term in the right-hand side of Eq. \eqref{eq:dsds} describes the non-linear transfer of circulicity from scale to scale. It is expressed as the divergence of a flux. The second term accounts for the unforced evolution of the mean circulicity $\rey{\ci} = \rey{s_is_i}/2$ and is scale-independent. 
The third one expresses the transfer of circulicity induced by viscous effects. 
The last one corresponds to the correlation between $\bs{s}$ and the external force $\bs{f}$ or more precisely its solenoidal component~$\bs{f}^s$.
Overall, the evolution equation of $\rey{\Delta s_i \Delta s_i}$ retains the same form as the equation of the second-order  structure-function of the velocity field classically derived in incompressible flows \cite{monin1975,frisch1995}.
However, the flux and dissipation terms of Eq.  \eqref{eq:dsds} are affected by the presence of density inhomogeneities and by compressibility.

More precisely, the non-linear transfer of circulicity involves three different components.
The first one, $\rey{\Delta v_j \Delta \jj_i\Delta \jj_i }$, is linked to the advection of $\bs{s}$ by the velocity field $\bs{v}$. It takes the form of a third-order structure-function. As shown in App. \ref{app:momentum}, its expression is identical to the one  appearing in the transfer flux of $\rey{\Delta \jj_i \Delta \jj_i}$. Hence, this first component is  representative of the transfer between scales not only of the solenoidal momentum, but also of the total momentum.
The remaining two contributions involve the increments of dilatational components of the momentum and velocity fields. 
Their presence is due to the fact that circulicity is driven by a production term $\pc$ and is not a conserved quantity. However, equation \eqref{eq:dsds} shows that the effects of this non-conservative  production can still be cast in the form of a conservative flux as far as the interscale transfer of circulicity is concerned.
Note that both of these  contributions vanish in constant-density turbulence. In that case, one is left with $\mathscr{F}_j = \rho^2 \rey{\Delta v_j \Delta v_i\Delta v_i}$, which is the flux obtained in the incompressible case weighted by $\rho^2$.

Concerning $\Einj$, it has two components. The first one is the mean dissipation of circulicity $\rey{\eps_c}$, the second  its mean production rate  $\rey{\pc}$ (see Eq. \eqref{eq:circ}).
The latter has the same origin as the two dilatational contributions of the non-linear transfer flux.
In this regard, a balance between these two dilatational terms and $\rey{\pc}$ would exist in the limit $r\to0$ if the turbulent fields were smooth. For rough fields, this is not case.
Note also that for constant-density turbulence, ${\pc}$ vanishes and one is left with $\Einj(t) = \mu \, \rho  \rey{\omega_i \omega_i}$, which is the solenoidal dissipation weighted by the density.
We would also like to emphasize that the definition of $\Einj$ stems from the unforced evolution of $\rey{\ci}$, as given by the first equality of Eq. \eqref{eq:EE}.
This origin is important because in forced stationary turbulence, one gets:
\begin{align} \label{eq:stationary}
\Einj = \rey{s_if_i^s}
\period
\end{align}
In that case, the value of $\Einj$ corresponds to the injection rate of circulicity. It is then set by the solenoidal component of the force and not by the total force. The implications of the latter property will be commented further in Sec.~\ref{sec:cascade}.

%-----------------------------------------------------------------------
\subsection{Monin-Yaglom relation} \label{sec:my}
%-----------------------------------------------------------------------
We now consider the limit $\mu \to 0$ and $r \to 0$ of the Karman-Howarth equation \eqref{eq:dsds}. The notation $\mu \to 0$ is used as a shorthand for referring to the high Reynolds number limit.
Then, provided the  correlation of the force $\bs{f}$ falls off exponentially, the only two terms that do not vanish in this limit are the non-linear flux divergence and $\Einj$.
Thus, in the limit $\mu \to 0$, $r\to0$, we obtain:
\begin{align} \label{eq:my}
\partial_{r_j} \left( \rey{\Delta v_j \Delta \jj_i\Delta\jj_i}  - 2  \rey{\Delta v_j^d \halfp{\jj_i\jj_i}}- 4 \rey{\Delta {d}_i  \halfp{\rho  v_i v_j}} \right)    = - 4 \Einj
\period
\end{align}
This equation is one of the main results of this work. It constitutes the Monin–Yaglom relation derived from  the solenoidal momentum evolution Eq. \eqref{eq:sol_mom}. 
Interestingly, this relation can be written in several alternative forms that may help shed light on different aspects of the transfer of circulicity.
For instance, using the Helmholtz decomposition \eqref{eq:q_decomp}-\eqref{eq:q_pot} and the Poisson equation \eqref{eq:poisson}, one may get rid of the dilatational components $v_i^d$ and $d_i$. Equation \eqref{eq:my} can indeed be rewritten as:
\begin{align} \label{eq:my2}
\partial_{r_j} \left( \rey{\Delta v_j \Delta \jj_i\Delta\jj_i}  - 2  \rey{\Delta v_j \halfp{\jj_i\jj_i}} + 4 \rey{\Delta {\jj}_j  \halfp{\pis}} \right)    = - 4 \Einj
\period
\end{align}
In agreement with this formulation, one may also rewrite $\Einj$ as $\Einj = \rey{\eps_c} + \frac{1}{2} \rey{\jj_i\jj_i\partial_j v_j} - \rey{\pis \partial_j (\rho v_j)}$.
Another possibility is to forgo the flux divergence formulation and allow for the presence of a source term. In particular, one may reformulate Eq. \eqref{eq:my} as:
\begin{align} \label{eq:my3}
\partial_{r_j}  \Big(\rey{\Delta v_j \Delta \jj_i\Delta\jj_i}\Big)  + \rey{\Delta \Theta \Delta (\jj_i\jj_i)} - 2 \rey{\Delta \Xi \Delta \pis} = - 4 \rey{\eps_c}
\coma
\end{align}
where we recall that $\Theta=\partial_j v_j$ and $\Xi = \partial_j(\rho v_j)$. This formulation is closer to the ones derived for instance in \cite{galtier2011,banerjee2014}. As in these references, the source terms can be associated with the dissipation term to provide a scale dependent effective dissipation $\eps_\mathrm{eff}$.

%-----------------------------------------------------------------------
\subsection{Circulicity cascade} \label{sec:cascade}
%-----------------------------------------------------------------------
The main physical interpretation of Eq. \eqref{eq:my}, and its variants Eqs. \eqref{eq:my2} and \eqref{eq:my3}, is similar to the one put forward in the classical incompressible case, except for two important points. First, it applies to circulicity and not energy, and second, the flux $\bs{\mathscr{F}}$ in Eq. \eqref{eq:my} is an ``effective'' flux that accounts not only for the scale-to-scale transfer of circulicity but also for its production by the dilatational source term $\pc$. The two processes are actually intertwined and setting a precise separation between the two bears some degree of arbitrariness. For this reason, the fact that these two effects can be collected within a single flux entity allows  to get rid of this distinction and to put forward a simple interpretation of the Monin–Yaglom equation \eqref{eq:my}. 
More precisely, the global flux $\F$ of circulicity flowing from scales larger than $r$ to scales smaller than $r$ is defined by  \cite{podesta2007}:
$$
\F = - \frac{1}{4 V_r} \int_{|\bs{r'}|\le r} \partial_{r_j} \mathscr{F}_j (\R') \ud \R' = - \frac{3}{4 r} \mean{\mathscr{F}_\prll}_{4\pi}(r)
\coma
$$
where the subscript $\prll$ denotes the coordinate of a vector in the direction $\hat{\R} = {\bs{r}}/{r}$ and where $\mean{\cdot}_{4\pi}$ refers to the average over the surface of the unit sphere:
$
\mathscr{F}_\prll = \mathscr{F}_j \hat{r}_j \andd
\mean{ \cdot }_{4\pi} = \frac{1}{4 \pi} \oint \cdot \ud  \hat{\R}
$.
From the definition of $\bs{\mathscr{F}}$, one can express this flux as:
\begin{align}
\F =  -\frac{3}{4r} \mean{ \rey{\Delta v_\prll \Delta \jj_i\Delta\jj_i}  - 2  \rey{\Delta v_\prll^d \halfp{\jj_i\jj_i}}- 4 \rey{\Delta {d}_i  \halfp{\rho  v_i v_\prll}} }_{4\pi}
\period
\end{align}
In isotropic turbulence, the brackets $\mean{\cdot}_{4\pi}$ can be dropped from this expression.

The Monin-Yaglom relation Eq. \eqref{eq:my} states that the global flux of circulicity is constant and equal to $\Einj$:
\begin{align} \label{eq:cascade}
\F = \Einj
\period
\end{align}
Alternatively, one can write that:
$$
\mean{\mathscr{F}_\prll}_{4\pi} = \mean{ \rey{\Delta v_\prll \Delta \jj_i\Delta\jj_i}  - 2  \rey{\Delta v_\prll^d \halfp{\jj_i\jj_i}}- 4 \rey{\Delta {d}_i  \halfp{\rho  v_i v_\prll}} }_{4\pi}  = - \frac{4}{3} \Einj r
\period
$$

Now, let us assume that the flow is forced and stationary. Then, the value of $\Einj$ is determined by the injection of solenoidal momentum: equation \eqref{eq:stationary} is valid and one has  $\Einj = \rey{s_i f_i^s} \ge 0$.
From there, two situations can be distinguished. 
\begin{itemize}
\item The first one occurs  when the solenoidal component of the force is not null. Then, we have :
$$
\Pi = \Einj>0 \period
$$
This condition indicates the existence of an effective circulicity cascade:  circulicity is transferred on average from large to small scales at a rate $\Einj$ independent from the scale $r$. This description is similar to the forward energy cascade associated with the Kolmogorov-Obukhov theory in incompressible turbulence.
\item The second situation arises when $\bs{f^s}=0$ and turbulence is only forced on dilatational modes. Then, the flux of circulicity becomes null: 
$$
\Pi = \Einj = 0
\period
$$
In that case, there is no effective circulicity cascade in the inertial range.
Instead, an effective non-linear equilibrium \cite{alexakis2018}  takes place in the inertial range, set by an exchange between the variances of the dilatational and solenoidal components of the momentum and constrained by the condition $\F=0$.
\end{itemize}
Thus, the existence of a circulicity cascade is not systematic and depends on the way turbulence is forced at large scales.
A parallel can be drawn between this conclusion and several observations which have been made in simulations of forced compressible turbulence \cite{federrath2010,federrath2013,wang2018,donzis2020,john2021}. 
Depending on whether the forcing is solenoidal or dilatational, significant differences were observed in the statistics of the simulated flow.
Even though the force is massic in \cite{federrath2010,federrath2013,wang2018,donzis2020,john2021} and not volumetric as in this study, a link between these distinct behaviors and the disappearance of the circulicity cascade would be worth investigating.

We close this section as we started it, by stressing that circulicity is not a conserved quantity and that the notions of its cascade and equilibrium state are, to some extent, ill-posed. This is why we used the term ``effective'' to qualify them. The flux $\bs{\mathscr{F}}$ and its integral $\Pi$ combine non-linear effects coming from conservative and non-conservative processes. 
In section \ref{sec:coarse}, we will use the coarse-graining approach to help disentangle the roles of cascading and production effects for circulicity in the inertial range.

%-----------------------------------------------------------------------
\subsection{Comparison with previous relations and practical use of Eq. \eqref{eq:my}}
%-----------------------------------------------------------------------
The Monin–Yaglom equation \eqref{eq:my}, and its variants \eqref{eq:my2}-\eqref{eq:my3}, are valid for compressible turbulence, whether in a highly compressible or weakly pseudo-compressible regime. As opposed to other such relations, for instance those derived in \cite{falkovich2010,galtier2011,wagner2012,banerjee2013,kritsuk2013,banerjee2014,banerjee2017,banerjee2018,andres2018,andres2019,ferrand2020,simon2021}, Eq. \eqref{eq:my} does not involve the actual pressure field and does not make any assumption on the equation of state, i.e. there is no hypothesis on whether the flow is isothermal, isentropic or polytropic.
What is more, all the terms appearing in Eq. \eqref{eq:my} can be evaluated knowing only the density $\rho$ and the velocity field $\bs{v}$ at a given time.
In this restricted sense, it is entirely decoupled from the energetic and mixing content of the flow.
Besides, it is also worth stressing that, in Eq. \eqref{eq:my}, the transfer between scales is written as the divergence of a flux $\bs{\mathscr{F}}$. 
By contrast, in the relations proposed in \cite{falkovich2010,galtier2011,wagner2012,banerjee2013,kritsuk2013,banerjee2014,banerjee2017,banerjee2018,andres2018,andres2019,ferrand2020,simon2021}, source terms are present which might not be easily recast in the form of a flux.

Compared to previous Monin–Yaglom relations there is also a major difference.
Most of these relations have been derived with a practical goal in mind: estimating the total injection and dissipation rates of energy. The relation we derive cannot fulfill the same goal.
Indeed, it can only provide information about the injection rate of solenoidal momentum $\Einj = \rey{\jj_i f_i^s} = \rey{s_i f_i^s}$.
Thus, from a practical point of view, the Monin–Yaglom equation Eq. \eqref{eq:my} cannot work as a substitute to the previously derived relations \cite{falkovich2010,galtier2011,wagner2012,banerjee2013,kritsuk2013,banerjee2014,banerjee2017,banerjee2018,andres2018,andres2019,ferrand2020,simon2021}.
Instead,  Eq. \eqref{eq:my} should be viewed as providing a complementary information. When coupled with previous relations, it should help disentangle the solenoidal and dilatational  contributions of the dissipation and energy injection.

%-----------------------------------------------------------------------
\subsection{Approximate Monin-Yaglom relation  in the weakly dilatational regime} \label{sec:approximate}
%-----------------------------------------------------------------------
In the Monin-Yaglom equation \eqref{eq:my}, the different contributions of the flux $\bs{\mathscr{F}}$ are respectively on the order of:
$$
\rey{\Delta v_j \Delta \jj_i\Delta\jj_i}
=
\mathcal{O}\left( \delta v \, \delta( \rho v) \,\delta( \rho v) \right)
\coma
\rey{\Delta v_j^d \halfp{\jj_i\jj_i}} =
\mathcal{O}\left( \delta v^d \, \delta( \rho^2 v^2) \right)
\andd
\rey{\Delta {d}_i  \halfp{\rho  v_i v_j}}  = \mathcal{O}\left( \delta d  \, \delta( \rho v^2) \right)
\coma
$$
where the notation $\delta X(r)$ refers to the order of magnitude of the increment $\Delta X$.
Provided $\delta \rho/\rho \to 0$ when $r\to 0$, one may approximate $\delta(\rho v)$ by $\rho \delta v$,  $\delta(\rho^2 v^2)$ by $\rho^2 \delta(v^2)$ and $\delta(\rho v^2)$ by $\rho \delta(v^2)$.
Therefore, the two dilatational contributions $\rey{\Delta v_j^d \halfp{\jj_i\jj_i}}$ and $\rey{\Delta {d}_i  \halfp{\rho  v_i v_j}}$ become negligible compared to the first one whenever:
\begin{align} \label{eq:weak_cond}
\delta v^d \ll \delta v \frac{(\delta v)^2}{\delta(v^2)} 
\andd
\delta d \ll \rho \delta v  \frac{(\delta v)^2}{\delta(v^2)} 
\period
\end{align}
These conditions define a particular regime that we call weakly dilatational.
We note that when $\bs{v}$ is solenoidal, $\delta(v^2)$ should be on the order of $(\delta v)^2$, as discussed in \cite{eyink_cours}. Therefore, to reach the weakly dilatational regime, it could  possibly be sufficient to verify :
\begin{align}\label{eq:weakdi}
\delta v^d \ll \delta v \andd \delta d \ll  \delta \jj
\period
\end{align}
An important point is that  conditions \eqref{eq:weak_cond} or \eqref{eq:weakdi}  only need to be verified at small scales: large scales can still possess large dilatational components. 
Besides, conditions \eqref{eq:weak_cond} and \eqref{eq:weakdi} are verified whenever $\delta v^d$ and  $\delta d$ decrease sufficiently fast when $r \to 0$, in particular when they decrease faster than their total counterparts. In that case, the inertial range may be divided in two ranges by a  limit scale $r_s$: below $r_s$,  conditions \eqref{eq:weak_cond} or \eqref{eq:weakdi} are satisfied while above $r_s$ they are not.

Simulations of subsonic and transsonic turbulence suggest that $\delta v^d \ll \delta v$ on most of the inertial range \cite{wang2018}, with $\delta v^d \sim r^{1/2}$ and $\delta v \sim r^{1/3}$. Simulations of supersonic turbulence \cite{federrath2010,ferrand2020,federrath2021} may be compatible with the idea that $\delta v^d \ll \delta v$ but only below the sonic scale. We are not aware of simulation results showing the behaviour of  the solenoidal and dilatational components of the momentum.

In any case, whenever the weakly dilatational regime defined by Eq. \eqref{eq:weak_cond} and possibly by Eq. \eqref{eq:weakdi} is reached,  the Monin-Yaglom equation \eqref{eq:my} can be simplified as:
\begin{align} \label{eq:my4}
\partial_{r_j}  \rey{\Delta v_j \Delta \jj_i\Delta\jj_i} =\partial_{r_j}  \rey{\Delta v_j^s \Delta s_i\Delta s_i} = - 4 \Einj
\period
\end{align}
This relation possibly applies to subsonic and transsonic turbulence and below the sonic scale for supersonic turbulence.

~\\ \indent
Another approximation of Eq. \eqref{eq:my} can also be formulated, always based on the idea that dilatational contributions can be neglected.
In \cite{aluie2011,aluie2012}, Aluie discarded the correlation between the pressure and velocity divergence by assuming that their co-spectrum decays as  $k^{-\beta}$ with $\beta > 1$. In terms of increments, Aluie made the assumption that $\rey{\Delta p \Delta \Theta} \propto r^{(\beta-1)/2}$, so that $\rey{\Delta p \Delta \Theta} \to 0$ when $r\to0$.
The same hypothesis can also be applied to the version \eqref{eq:my3}  of the present Monin-Yaglom relation. 
As in \cite{aluie2011,aluie2012}, we can assume that the dilatational source terms $\rey{\Delta \Theta \Delta (\jj_i\jj_i)}$ and $\rey{\Delta \Xi \Delta \pis}$ go to $0$ when $r \to 0$. 
In that case, the Monin-Yaglom relation \eqref{eq:my3} is approximated by:
\begin{align} \label{eq:my5}
\partial_{r_j}  \rey{\Delta v_j \Delta \jj_i\Delta\jj_i} = - 4 \rey{\eps_c}
\period
\end{align}
The two approximations \eqref{eq:my4} and \eqref{eq:my5} coincide provided $\Einj = \rey{\eps_c}$. This occurs when $\rey{\pc} \ll \rey{\eps_c}$. This condition can be met when dilatational motions are negligible not only at small scales but also on the whole spectrum and in particular, when the dilatational forcing is weak.
In that case, the circulicity injected at large scales is equal to the circulicity dissipated at small scales. Equations \eqref{eq:my4} and \eqref{eq:my5} are then indicative of a circulicity cascade.

%-----------------------------------------------------------------------
\subsection{Inertial range scalings in the weakly dilatational regime}
%-----------------------------------------------------------------------
In incompressible turbulence, the classical Monin-Yaglom relation can be used to determine the scaling of velocity increments \cite{frisch1995}. To this end, it is assumed that velocity increments obey a self-similar scaling such that $\Delta v_i \stackrel{\text{law}}{\propto}  r^{n_v}$. In particular, the velocity structure function of order $p$ verifies:
\begin{align} \label{eq:struct}
\rey{|\Delta \bs{v}|^p} \propto r^{n_v p}
\period
\end{align}
The incompressible Monin-Yaglom relation then constrains $n_v$ to be equal to $1/3$.
It also implies that the prefactor of equation \eqref{eq:struct} is the solenoidal dissipation $\epsr^{1/3}$.
These predictions must of course be corrected to account for internal intermittency when necessary \cite{frisch1995}.

In Sec. \ref{sec:approximate}, we argued that below a sonic-like scale, under conditions \eqref{eq:weakdi}, the Monin–Yaglom relation \eqref{eq:my} could be approximated by Eq. \eqref{eq:my4}.
In addition to this approximation, we assume that the increments of  the different flow variables obey a self-similar scaling as in incompressible turbulence. For a given variable $X$, we write:
\begin{align}\label{eq:selfsim}
 \Delta X \stackrel{\text{law}}{\propto} r^{n_X}
\end{align}
where $n_X$ is the power law exponent associated with the variable $X$.
Then, the Monin-Yaglom relation \eqref{eq:my4} forces the relation:
$$
n_v + 2 n_j = {1} 
\period
$$
Furthermore,  we conjecture that $\Delta \rho \to 0$ as $r\to 0$ and more specifically that $n_\rho \ge n_v$. Then, using the relation $\Delta (\rho X) = \Delta \rho \halfp{X} + \halfp{\rho} \Delta X$, one deduces that $n_{\jj} =  n_v$. Therefore, with this set of assumptions, one arrives at the conclusion that:
$$
n_v = n_j = \frac{1}{3}
\period
$$
This is the usual Kolmogorov-Obukhov scaling of the velocity increment, corresponding to a $k^{-5/3}$ spectrum.
Note that the self-similarity assumption \eqref{eq:selfsim} does not allow to separate the exponents of the different components of the velocity and momentum fields. Given the weak dilatational assumption, which is summed by Eq. \eqref{eq:weakdi}, the common value of $n_j$ and $n_v$ is that of the principal components of these fields, it is to say, their solenoidal components. The dilatational components can have a different exponent as long as it is larger than $1/3$.

There is also another important conclusion that can be derived from the approximate Monin-Yaglom relation in the weakly dilatational regime.
The sole dimensioning parameter that appears in this law is $\rey{\eps_c}$, which is a  dissipation rate based on solenoidal fields and on the sole shear viscosity $\mu$.
Therefore, one can refine the similarity assumption for the velocity and momentum fields and write that:
\begin{align}
\Delta v_i \stackrel{\text{law}}{\propto} \left( (\rey{\eps_c }/\rh^2) \, r \right)^{1/3}
\andd
\Delta \jj_i \stackrel{\text{law}}{\propto} \left( (\rh \; \rey{\eps_c }) \, r \right)^{1/3}
\end{align}
This prediction is reminiscent of  several simulations where turbulent velocity spectra with a Kolmogorov-Obukhov scaling are seen to depend on the solenoidal dissipation rate of energy and not the total one \cite{wang2018,john2019}.

%-----------------------------------------------------------------------
\subsection{Inertial range scalings with shocks}
%-----------------------------------------------------------------------
 When shocks are present in the flow, the self-similarity assumption \eqref{eq:selfsim} cannot be upheld. In particular,  Lindborg \cite{lindborg2019} showed that when dissipation occurs exclusively in shocks, the velocity field does not verify equation \eqref{eq:struct}. Instead, one has:
\begin{align} \label{eq:struct2}
\rey{|\Delta \bs{v}|^p} = \mean{|\Delta \bs{v}|^p}_\mathtt{sh} \frac{r}{d_\mathtt{sh}}
\coma
\end{align}
where $d_\mathtt{sh}$ is the mean distance between shocks and $\mean{|\Delta \bs{v}|^p}_\mathtt{sh}$ is the average jump of $|\Delta \bs{v}|^p$ across shocks.

If we want to analyze scalings compatible with Eq. \eqref{eq:my} we must consequently account for these two distinct behaviours.
To this end, 
we propose to separate the average of two-point correlations in order to make shock contributions explicit. For a given quantity $X(\x,\x')$ depending on the two positions $\x$ and $\x'=\x+\R$, we split its average as:
\begin{align}
\rey{X}(\R)  = \Ps(\R)  \msh{X}(\R) + (1-\Ps(\R))\mns{X}(\R)
\coma
\end{align}
where $P_\mathtt{sh}$ is the probability of finding a shock between two points separated by $\R$ and where $\msh{X}$ and $\mns{X}$ are the averages of $X$ conditioned respectively on the event of finding or not a shock between two points separated by $\R$.

Then, the Monin-Yaglom relation \eqref{eq:my} can be written as:
\begin{align} \label{eq:my_sh}
 & \partial_{r_j} \left(  \Ps \left( \msh{\Delta v_j \Delta (\rho v_i) \Delta (\rho v_i)}  - 2  \msh{\Delta v_j \halfp{\rho^2 v_i v_i}} - 4 \msh{\Delta d_i \halfp{\rho v_iv_j}} \right) \right)
\nonumber \\ 
+ &
 \partial_{r_j} \left(  (1-\Ps) \left( \mns{\Delta v_j \Delta (\rho v_i) \Delta (\rho v_i)}  - 2  \mns{\Delta v_j \halfp{\rho^2 v_i v_i}} - 4 \mns{\Delta d_i \halfp{\rho v_iv_j}} 
\right)  \right)    = - 4 \Einj
\end{align}
This equation only makes explicit the presence of shocks but does not introduce any additional assumption compared to the Monin-Yaglom relation \eqref{eq:my} from which it is derived.
However, if we want to  interpret it further,  some simplifications must be made.
First, for any given quantity $X(\x,\x')$ with $\x$ and $\x'=\x+\R$, we assume that $\msh{X}$ is dominated by the shock contribution. In that case, we can write that:
$$
\msh{X}(\R) = \mean{X}_\mathtt{sh}
\coma
$$
where $\mean{X}_\mathtt{sh}$ represents the contribution of the shock jumps to $X$, averaged over all shocks. This contribution is independent from $\R$.
Second, we assume that non-shocked contribution to increments obeys a power law. Thus, we reduce Eq. \eqref{eq:selfsim} to the non-shocked intervals:
$$
\left. \Delta X\right|_\mathtt{ns} \stackrel{\text{law}}{\propto}  r^{n_X}
$$
Finally, we consider that the spatial repartition of shocks  follows an isotropic Poisson distribution with a mean distance $d$. As a result, as in \cite{lindborg2019}, one has $\Ps \approx \frac{r}{d}$ for $r\ll d$ and $1-\Ps \approx 1$.
With these approximations, the Monin-Yaglom relation \eqref{eq:my_sh} becomes:
\begin{align} \label{eq:my_sh2}
 &
 \partial_{r_j} \left(   \mns{\Delta v_j \Delta (\rho v_i) \Delta (\rho v_i)}  - 2  \mns{\Delta v_j \halfp{\rho^2 v_i v_i}} - 4 \mns{\Delta d_i \halfp{\rho v_iv_j}} 
\right)      = - 4 ( \Einj - \mathcal{E}_\mathtt{sh})
\\ \nonumber
\with & 
\mathcal{E}_\mathtt{sh} = -   \frac{1}{4d_\mathtt{sh}} \left( \mean{\Delta v_\parallel \Delta (\rho v_i) \Delta (\rho v_i)}_\mathtt{sh}  - 2  \mean{\Delta v_\parallel \halfp{\rho^2 v_i v_i}}_\mathtt{sh} - 4 \mean{\Delta d_i \halfp{\rho v_iv_\parallel}}_\mathtt{sh} \right)
\end{align}
If we except the particular case $\Einj=\mathcal{E}_\mathtt{sh}$, this formula can only be fulfilled if at least one of the terms in the divergence on the left-hand side scales as $r$. Given the scaling laws assumed in the non-shocked region, this condition implies that at least one of this condition must be verified:
\begin{align}\label{eq:ineq}
n_v + 2 n_{\rho v} \le 1 \coma
n_v + n_{\rho^2 v^2} \le 1 \coma
n_d + n_{\rho v^2} \le 1
\period
\end{align}
We note that, since $\rey{\Delta (\rho v_i) \Delta (\rho v_i)} = \rey{\Delta s_i \Delta s_i} +  \rey{\Delta d_i \Delta d_i}$, one must have $n_d \ge n_{\rho v}$.
Furthermore, if we assume that $\Delta \rho \to 0$ as $r\to 0$ and more specifically that $n_\rho \ge n_v$, then, using the relation $\Delta( \rho X) = \Delta \rho \halfp{X} + \halfp{\rho} \Delta X$, one can show that $n_{\rho v} =  n_v$, $n_{\rho^2 v^2} = n_{v^2}$ and $n_{\rho v^2} = n_{v^2}$. Finally, the Schwartz relation $\rey{\Delta (v_iv_i)^2} \le 2 \sqrt{\rey{(\halfp{v_j}\halfp{v_j})^2}} \sqrt{\rey{(\Delta v_i\Delta v_i)^2}}$ implies that $n_{v^2} \ge n_v$.

From there, we finally arrive at the conclusion that the first inequality of equation \eqref{eq:ineq} can be verified if $n_v \le 1/3$, while the two others require $n_v \le 1/2$. This can be summed up as:
\begin{subequations}
\begin{align}
\text{Non-vanishing } \mns{\Delta v_j \Delta (\rho v_i) \Delta (\rho v_i)} ~~~~~ &: ~~~~ n_v \le \frac{1}{3}
\coma
\\
\text{Non-vanishing } - 2  \mns{\Delta v_j \halfp{\rho^2 v_i v_i}} + 4 \mns{\Delta d_i \halfp{\rho v_iv_j}} ~~~~ &: ~~~~
n_v \le \frac{1}{2}
\period
\end{align}
\end{subequations}
The upper bound of the first constraint corresponds to the usual Kolmogorov-Obukhov scaling of the velocity increment, corresponding to a $k^{-5/3}$ spectrum.
The upper bound of the second corresponds to a velocity spectrum scaling as $k^{-2}$.
The first scaling is traditionally observed in simulations with moderate Mach numbers and with solenoidal forcing \cite{wang2018}.
By contrast, in simulations of supersonic turbulence, a $k^{-2}$ spectrum is obtained for scales larger than the sonic scale.  Below this scale, a shallower spectrum with a scaling close to $k^{-5/3}$ is observed \cite{ferrand2020,federrath2021}.
We recall that these scalings are derived for the non-shocked regions of the flow.
Given the decomposition $\rey{\Delta v_i \Delta v_i} = \Ps \msh{\Delta v_i \Delta v_i} + (1-\Ps) \mns{\Delta v_i \Delta v_i}$, we note that the $k^{-2}$ spectrum observed in simulations of supersonic turbulence can have two contributions. The first comes from the presence of shocks, as in Burgers turbulence. The other comes from the dilatational components of the velocity and momentum, shock excluded. The role of this second contribution was also highlighted in \cite{ferrand2020}.

To conclude on this topic, we would like to recall that without solenoidal forcing, when $\Einj = \rey{\jj_i f_i^s}=0$, an equilibrium takes place in the inertial range. Equation \eqref{eq:my_sh2} shows that this equilibrium can be understood as a balance between the shocked and non-shocked regions of the flow.
In particular, if $\mathcal{E}_\mathtt{sh} > 0$, shocks lead to a forward transfer of circulicity from large to small scales, while a backward transfer occurs for non-shocked regions.

%=======================================================================
\section{Coarse-graining approach} \label{sec:coarse} 
%=======================================================================
In this section, we return to discussing the existence of a circulicity cascade. However, instead of the point-splitting methodology which led us to derive the Monin-Yaglom equation \eqref{eq:my}, we now use the coarse-graining approach which was initially propelled by Onsager's work \cite{onsager1949,eyink2006}.
This approach does not require the flow to be homogeneous, and unless said otherwise, we drop this hypothesis.
As a whole,  we closely follow the works of Eyink \& Drivas and Aluie  \cite{aluie2011,aluie2012,aluie2013,eyink2018}. 

%-----------------------------------------------------------------------
\subsection{Resolved and subscale circulicities}
%-----------------------------------------------------------------------
The basic tool of the coarse-graining approach is spatial filtering.
For a given variable $X(\x,t)$, one defines its filtered value by:
\begin{align}
\flt{X}(\x,t) = \int G_\ell (\R) X(\x + \R,t) \ud \R
\coma
\end{align}
where $G_\ell(\R) = G(\R/\ell)\ell^{-3}$ is a filter function with a kernel $G$ rapidly decreasing at infinity and verifying the normalization condition $\int G(\R) \ud \R = 1$.
The idea is then to look at the properties of the filtered circulicity $\flt{\ci}$.
More precisely,  $\flt{\ci}$ can be split into a resolved and subscale circulicity:
$$
\flt{\ci}  = \res{\ci} +\sub{\ci}  \with  \res{\ci} = \frac{1}{2} \flt{s_i}\flt{s_i}  \andd  \sub{\ci} = \frac{1}{2} \tau(s_i, s_i)
\coma
$$
where we used the notation $\tau(X,Y)$ to denote the subscale moment of given quantities $X$ and $Y$:
$$
\tau(X,Y) = \flt{XY} - \flt{X} \flt{Y}
\period
$$
The evolution of the resolved circulicity $\res{\ci}$ can be deduced by filtering the evolution equation  \eqref{eq:ns_qs} of $\bs{s}$ and multiplying it by $\flt{\bs{s}}$. As for the subscale circulicity, its evolution can be derived by filtering the circulicity equation \eqref{eq:circ} and substracting the resolved part.
By doing so, we obtain that:
\begin{subequations} 
\begin{align}
\label{eq:res_circ}
  \partial_t \res{\ci} + \partial_j \res{\mathcal{F}_j^c}
 =&
- \Pi_\ell - \pi^{f\mu}_\ell  + \res{\pc}+ \flt{s_i f_i^s}
\coma
 \\
\label{eq:sub_circ}
 \partial_t \sub{\ci} + \partial_j \sub{\mathcal{F}_j^c}
= &
\phantom{ - }~ \Pi_\ell + \pi^{f\mu}_\ell + \sub{\pc}- \flt{\eps_c}  
\period
\end{align}
\end{subequations}
In these equations, $\res{\bs{\mathcal{F}^c}}$ and $\sub{\bs{\mathcal{F}^c}}$ are the resolved and subscale circulicity fluxes. 
We recall that the local circulicity flux appearing in Eq. \eqref{eq:circ} is noted $\bs{\mathcal{F}^c}$. If we mark symbolically the dependency of this flux upon the flow variables as $\bs{\mathcal{F}^c}\equiv\bs{\mathcal{F}^c}[\bs{X}] $, then we have : $\res{\mathcal{F}_j^c} = \mathcal{F}^c_j[\flt{\bs{X}}] + \flt{\jj_i} \tau(\jj_i,v_j)$ and $\sub{\mathcal{F}_j^c} = \flt{\mathcal{F}^c_j[{\bs{X}}]} - \mathcal{F}^c_j[\flt{\bs{X}}] - \flt{\jj_i} \tau(\jj_i,v_j)$. These two fluxes do not play any role in the forthcoming analysis.
The term $\pi_\ell^{f\mu}$  combines a filtered scale dissipation and the subscale contribution of the force $\bs{f}^s$. It is defined as $\pi_\ell^{f\mu} = \tau(s_i,fi^s) + \mu \flt{\Omega_i} \flt{\omega_i}$. Like the fluxes, the role of $\pi_\ell^{f\mu}$ is inessential since it vanishes in the high Reynolds limit and with the assumption that the forcing is at large scale.
The last three terms that need to be defined,  $\res{\pc}$, $\sub{\pc}$ and $\Pi_\ell$, are the ones playing a central role in the present analysis. 
The first two correspond to the resolved and subscale contributions of the circulicity production term $\pc$. More precisely, we have: 
\begin{subequations} \label{eq:pc_decomp}
\begin{align}
& \flt{\pc} = \res{\pc} + \sub{\pc}
\\
 \text{with}~~&  
\res{\pc} = - \frac{1}{2} | \flt{\jj_i}\flt{\jj_i} |^2 \flt{\Theta}  + \flt{\pis} \flt{\Xi}
\\ 
\text{ and }~~&
\sub{\pc} = - \frac{1}{2} \left(  \tau(j_ij_i, \Theta) + \flt{\Theta} \tau(\jj_i,\jj_i) \right)
+ \tau(\pis,\Xi)
\period
\end{align}
\end{subequations}
The third term is defined by:
\begin{align}\label{eq:pil}
\Pi_\ell = - \tau(v_j,\jj_i) \partial_j \flt{\jj_i} 
\period
\end{align}
This term consists in a subscale-stress multiplying the gradient of a filtered quantity and marks the interaction between resolved and unresolved scales.  
It appears as a sink  in the resolved circulicity equation \eqref{eq:res_circ} and as a source in the subscale circulicity equation \eqref{eq:sub_circ}.
Thus, $\Pi_\ell$ can be interpreted as a non-linear exchange term between the resolved and subscale circulicities. 

Note that if we introduce a density-weighted filter and density-weighted subscale moments as $\fflt{X} = \flt{\rho X}/\flt{\rho}$ and $\tilde{\tau}(X,Y) = \fflt{XY} - \fflt{X}\fflt{Y}$, then the exchange term $\Pi_\ell$ can also be written as:
\begin{align} \label{eq:pil2}
\Pi_\ell = - \flt{\rho}^2 \tilde{\tau}(v_i,v_j) \,\partial_j \fflt{v_i} - \tau(\rho,v_j) \partial_j\big(\flt{\rho} \,\fflt{v_i}\fflt{v_i}\big) - \tau(v_j,\jj_i) \fflt{v_i}\partial_j\flt{\rho}
\period
\end{align}
Up to a density factor, the first term on the right-hand side is the deformation work that plays a central role in the transfer of kinetic energy between scales \cite{aluie2011,aluie2012,aluie2013,eyink2018}.
Besides, the second term is similar to the ``baropycnal work'' which also appears as a means to transfer kinetic energy. However, the actual pressure $p$ which enters the definition of the ``baropycnal work'' is here replaced with a dynamic one: $\flt{\rho} \fflt{v_i}\fflt{v_i}$.
Finally, the last term has no direct equivalent in the transfer of kinetic energy.
This comparison between $\Pi_\ell$ and the transfer of kinetic energy reinforces the idea that $\Pi_\ell$ is an exchange term between resolved and unresolved scales. Furthermore, it indicates that some of the mechanisms  at work in the transfer of energy and circulicity are similar.

To follow on this comparison, we note that the exchange term between the resolved and subscale kinetic energies has been a staple of the analysis of subgrid-scale models for large-eddy simulations (LES) for more than five decades (see \cite{sagaut2006,garnier2009} and references therein). Its link with the Kolmogorov-Obukhov spectrum and with the Monin-Yaglom relation has long been discussed \cite{sagaut2006,garnier2009}.
However, its use as a proper theoretical tool able to demonstrate the existence of a cascading process seems to have been only recognised more recently, most notably with Refs. \cite{aluie2011,aluie2012,aluie2013,eyink2018}.

%-----------------------------------------------------------------------
\subsection{Circulicity cascade}
%-----------------------------------------------------------------------
We now consider the high Reynolds limit of the resolved and subscale circulicities, with the added assumption that the force only acts at large scales. Taking the limit  $\mu \to 0$ in  Eqs. \eqref{eq:res_circ} and \eqref{eq:sub_circ}, and discarding the subscale force contribution, we obtain that:
\begin{subequations} 
\begin{align}
\label{eq:res_lim}
  \partial_t \res{\ci} + \partial_j \res{\mathcal{F}_j} 
 =&
 -  \Pi_\ell^* - \sub{\pc}^*
+ \flt{\pc^*}
+ \flt{s_i f_i^s}
\coma
 \\
\label{eq:sub_lim}
 \partial_t \sub{\ci} + \partial_j \sub{\mathcal{F}_j} 
= &
\phantom{ - }~ \Pi_\ell^* + \sub{\pc}^*  - \flt{\eps_c^*}  
\period
\end{align}
\end{subequations}
where we substituted $\res{\pc}$ with $\flt{\pc} - \sub{\pc}$ and where:
$$
\Pi_\ell^* = \lim_{\mu \to 0} \Pi_\ell \coma \eps_c^* = \lim_{\mu \to  0} \eps_c
\coma
\sub{\pc}^* =  \lim_{\mu \to 0} \sub{\pc }
\andd
\pc^* = \lim_{\mu \to 0} \pc
\period
$$
These limits are taken in a weak distributional sense. For instance, one has:
$$
\pc^* = - \frac{1}{2} | \jjv|^2 * \Theta + \pis * \Xi
\coma
$$
where the operator  $*$  stands for the distributional product in the limit $\mu \to 0$.
The reason for introducing this distributional meaning is explained in \cite{eyink2018}. It  comes from the fact that $\rho$ and $\bs{v}$ become non-smooth when $\mu \to 0$. Hence, $\Theta=\partial_j u_j$ and $\Xi=\partial_j(\rho u_j)$ exhibit divergences that can only be described by distributions. 
As a result, products such as $|\jjv|^2 \Theta$ and $\pis  \Xi$ behave as the products of Heaviside functions with Dirac distributions. 
Their values are ill-posed in the sense that they depend on the particular regularizing path chosen to describe them \cite{dalmaso1995,pares2006}. 
This regularizing path  does not need to be made explicit for our purpose. It is sufficient to know that it depends on the physics of the problem and on the particular way the limit $\mu \to 0$ is approached.

To discuss the existence of a circulicity cascade, we make two crucial hypotheses, similar to those made in \cite{eyink2018}.
The first one is that dissipative anomalies exist and that the dissipation of circulicity $\eps_c$  has a non-zero finite value in the infinite Reynolds limit:
\begin{align}
\eps_c^*  \ne 0 
\period
\end{align}
The second is that in the limit $\ell \to 0$, the evolution equation \eqref{eq:res_lim} of $\res{\ci}$ tends,  in the sense of distributions,  to the evolution equation of $\ci$ in the ideal limit $\mu \to 0$.
The limit of Eq. \eqref{eq:res_lim} can be written as follows:
\begin{align} \label{eq:circ_lim2}
  \partial_t \ci + \partial_j \mathcal{F}_j^c 
 =&
\pc^*
+ s_i f_i^s
 - \Pi_0^* - \subz{\pc}^*
\coma
\end{align}
with $\Pi_0^* = \lim_{\ell\to0} \Pi_\ell^*$ and $\subz{\pc}^* = \lim_{\ell\to0}\sub{\pc}^*$.
 The evolution of $\ci$ in the ideal limit is deduced from Eq. \eqref{eq:circ} and can be written as:
\begin{align} \label{eq:circ_lim}
\partial_t \ci + \partial_j \mathcal{F}_j^c =    \pc^* + s_if_i^s - \eps_c^*  
\period
\end{align}
The reason for the identity between these two limit equations is discussed in \cite{eyink2018} (p. 10, second paragraph) and comes for the principle that ``objective physical facts such as the rate of decay of energy [...] cannot depend upon an arbitrary scale $\ell$''. One may expect that this ``objectivity principle'' also applies to circulicity, its dissipation and injection rates.
Comparing Eqs. \eqref{eq:circ_lim} and \eqref{eq:circ_lim2}, one deduces that: 
\begin{align} \label{eq:cascade2}
\Pi_0^* + \subz{\pc}^* = \lim_{\ell \to 0} \lim_{\mu \to 0} \Pi_\ell + \sub{\pc} = \eps_c^* 
\period
\end{align}
This equality is the main result of this section. 
It states that the circulicity transferred non-linearly from resolved to subscales ($\Pi_0^*$) summed to the one produced at subscales ($\subz{\pc}^*$) does not vanish in the limit $\mu\to 0, \ell \to 0$ : it tends to a non-zero finite value independent from $\ell$ and equal, in the sense of distributions, to the dissipation $\eps_c^*$.
Equation \eqref{eq:cascade2} does not assert the existence of a circulicity cascade. Instead, it expresses a balance between transfer, production and dissipation of circulicity in the subscale range. 
Since $\eps_c^*\ne 0$, this balance is anomalous and both the transfer and subscale production of circulicity may be the source of the anomalous dissipation $\eps_c^*$.
A similar result has already been obtained for the kinetic energy: in \cite{eyink2018}, it was shown that the anomalous dissipation of energy could have two origins, an energy cascade and a pressure-dilatation defect akin to a subscale energy production. Both elements are similar to the mechanisms identified in this work for circulicity.

Equation \eqref{eq:cascade2} has been cast in the form of a subscale balance. It can also be used to express a balance for resolved scales.
Indeed, using Eq. \eqref{eq:pc_decomp}, one can show that:
\begin{align} \label{eq:cascade3}
\resz{\pc}^* - \Pi_0^* = \lim_{\ell \to 0} \lim_{\mu \to 0}  \res{\pc} - \Pi_\ell  = \pc^* - \eps_c^* 
\end{align}
This equation shows that, in the limit $\mu\to0$, $\ell\to0$, the circulicity transferred to small scales and produced at resolved scaled tends to a finite value equal to the difference between the production and dissipation of circulicity.

Equations \eqref{eq:cascade2} and \eqref{eq:cascade3} can be understood as the coarse-grained equivalents of the Monin-Yaglom relations \eqref{eq:my3} and \eqref{eq:my}.
However, as opposed to a Monin-Yaglom relation, Eqs. \eqref{eq:cascade2} and \eqref{eq:cascade3} are not statistical  and do not involve an ensemble mean or, if ergodicity applies,  an integration over a spatial domain. Instead, they are written in a weak distributional sense. In this regard, it is worth stressing that equation \eqref{eq:cascade2} does not imply that $\Pi_0^* +  \subz{\pc}^* = \eps_c^*$ locally and instantly. As a weak equality, it involves spatial integrations with test functions. When seen under this light, equality \eqref{eq:cascade2} has a meaning which is indeed stronger that the one given to a Monin-Yaglom relation, but, at the same time, which is maybe not so far removed from it. 
In the same line of thought, it should also be mentioned that in a real flow, with finite $\mu$ and $\ell$, the measurable quantities which give rise to equality \eqref{eq:cascade2} are $\Pi_\ell$, $\sub{\pc}$ and $\flt{\eps_c}$ which appear in the right-hand side of Eq. \eqref{eq:sub_circ}. The left-hand side of this equation has no reason to be equal to $0$ so that the values of $\Pi_\ell +\sub{\pc}$ and $\flt{\eps_c}$ have no reason to be equal locally and instantaneously. Again, this is not what equation \eqref{eq:cascade2} is meant to say. The weak distributional sense of this equation should always be kept in mind in order to avoid  over-interpreting it. The same also applies to Eq. \eqref{eq:cascade3}.

Another remark worth making about equalities \eqref{eq:cascade2} and \eqref{eq:cascade3} comes from the fact that a double limit is taken, first $\mu \to 0$ and then $\ell \to 0$. In doing so, Eyink \& Drivas \cite{eyink2018} noted that several distributional products appear. For instance, the contribution $\tau(\pis,\Xi)$ of $\sub{\pc}$ is equal to $\flt{\pi \Xi} - \flt{\pi}\flt{\Xi}$. Taking the double limit, then leads to write:
$$
\lim_{\ell \to 0} \lim_{\mu \to 0} \tau(\pis,\Xi) = \pis * \Xi - \pis \circ \Xi 
\period
$$
In this equality, the symbol $\circ$ refers to the regularization of $\flt{\pi}\flt{\Xi}$ which is different from the one associated with the symbol $*$ and the regularization of $\flt{\pi \Xi}$. 

To conclude this discussion on Eqs. \eqref{eq:cascade2} and \eqref{eq:cascade3}, we would like to mention that a purely statistical version of these equalities can also be derived, without invoking the finiteness of $\eps_c^*$ and an ``objectivity principle''.
Instead of these assumptions, one may assume that the flow is homogeneous and statistically stationary, and that the mean of $\eps_c^*$ is finite, i.e. that $\lim_{\mu \to 0} \rey{\eps_c} \ne 0$, which is a weaker statement than $\eps_c^* \ne 0$. Then, by averaging Eqs. \eqref{eq:sub_circ} and \eqref{eq:res_circ}, one deduces that, in the limit $\mu \to 0$ and $\ell \to 0$, the following equality stands:
\begin{align} \label{eq:cascade4}
\lim_{\ell \to 0} \lim_{\mu \to 0} \rey{\Pi_\ell} + \rey{\sub{\pc}}= \rey{\eps_c}
\andd
\lim_{\ell \to 0} \lim_{\mu \to 0}   \rey{\Pi_\ell} - \rey{\res{\pc}}  = \rey{\eps_c} - \rey{\pc} = \Einj
\period
\end{align}
These two relations can be obtained by averaging Eqs. \eqref{eq:cascade2} and \eqref{eq:cascade3}. They can also be derived by filtering the Monin-Yaglom relations \eqref{eq:my3} and \eqref{eq:my}.
Conditions under which  $\rey{\sub{\pc}}$ and $\rey{\res{\pc}}$ could be neglected compared to $ \rey{\Pi_\ell}$ were given in Sec. \ref{sec:approximate}. They are met in a so-called weakly dilatational regime. In that case, the mean transfer of circulicity $\rey{\Pi_\ell}$ becomes independent of $\ell$ and tends to a finite non-zero value : a circulicity cascade takes place.
Otherwise, a mean transfer-production-dissipation equilibrium occurs in the inertial range in lieu of this sole cascade.
Still, in Sec. \ref{sec:cascade}, it was shown that this mean equilibrium could be interpreted as an effective cascade by recasting the source term in the form of a flux-like transfer term.

%-----------------------------------------------------------------------
\subsection{Increment scalings and locality}
%----------------------------------------------------------------------
There are two ways by which the anomalous dissipation $\eps_c^*$ can be sustained : one is the transfer term $\Pi_\ell$, the other is the subscale production $\sub{\pc}$.
The orders of magnitude of these terms are linked to the space increments of the velocity, density and momentum fields.
The order of magnitude of the increment $\Delta X(\x,\R)$ of a quantity $X$ is noted $\delta X(r)$.
As discussed in \cite{aluie2011,eyink2018,eyink_cours}, one has:
$$
\tau(X,Y) = \mathcal{O}(\delta X \delta Y) \coma
\partial_i \flt{X} = \mathcal{O}\left(\frac{\delta X}{\ell}\right) \coma
X - \flt{X} = \mathcal{O}(\delta X)
\period
$$
Using these relations, it is straightforward to show from Eq. \eqref{eq:pil} that:
$$
\Pi_\ell = \mathcal{O}\left( \frac{\delta v(\delta j)^2}{\ell}\right)
\period
$$
One may also use Eq. \eqref{eq:pil2} and the developments of Aluie \cite{aluie2011} to show that this expression can be simplified as:
\begin{align} \label{eq:pil_om}
\Pi_\ell =  \mathcal{O}\left( \frac{(\delta v)^3}{\ell}\right) \rho^2 \left[ 1 + \mathcal{O}\left(\frac{\delta \rho}{\rho}\right)+ \mathcal{O}\left(\frac{\delta \rho^2}{\rho^2}\right) \right]
\period
\end{align}
From this estimate, one deduces that the transfer term does not vanish in the limit $\ell \to 0$ provided the velocity field is rough enough. 
More precisely, let us assume that the order of magnitude $\delta X$ of the increment a quantity $X$ obeys a power law:
$$
|\delta X(\ell)^n| \propto \ell^{\zeta_n^X}
\period
$$
Then, if $\zeta_n^\rho > 0$, i.e. if $\delta \rho/\rho \to 0$ when $\ell \to 0$, Eq. \eqref{eq:pil_om} implies that $\Pi_\ell$ is non-vanishing provided:
\begin{align}\label{eq:casc}
\zeta_n^v \le \frac{n}{3} \for n \ge 3
\period
\end{align}
This is a necessary condition   for $\Pi_\ell$ to contribute to the anomalous dissipation $\eps_c^*$. This condition is identical to the one derived in \cite{eyink2018} for the energy cascade. This is not entirely unexpected: the resolved to unresolved exchange terms of circulicity and energy both involve the deformation work as a common mechanism.

We now turn our attention to the subscale circulicity production $\sub{\pc}$ defined in Eq. \eqref{eq:pc_decomp}.
This term has  three components involving $\Theta = \partial_j v_j = \partial_j v_j^d$ and $\Xi=\partial_j \jj_j = \partial_j d_j$.
By comparing the right- and left-hand sides of the relation $\partial_i \flt{\Theta} = \partial^2_{ij} \flt{v_j^d}$, one deduces that $\delta \Theta/\ell \sim \delta v^d/\ell^2 $ so that $\delta \Theta \sim \delta v^d/\ell$. 
Similarly, one derives that $\delta \Xi \sim \delta d/ \ell$.
Therefore, the three contributions of $\sub{\pc}$ lead to the following order of magnitude:
$$
\sub{\pc} = \mathcal{O}\left( \frac{\delta v^d \delta(j^2) }{\ell} \coma \frac{\delta v^d (\delta j)^2 }{\ell}   \coma \frac{\delta \pis \delta d}{\ell} \right)
\period
$$
The Poisson equation for the pressure \eqref{eq:poisson} also leads to $\delta \pis  \sim \delta (v \jj)$.
Thus, the conditions for the production to be non-vanishing are:
$$
\zeta_n^{v^d} + \zeta_n^{j^2} \le 1
\coma
\zeta_n^{v^d} + 2 \zeta_n^{j} \le 1
\coma
\zeta_n^{d} + \zeta_n^{vj} \le 1
\for 
n \ge 3
\period
$$
If the dilatational fields $\bs{v}^d$ and $\bs{d}$ decay sufficiently rapidly then these conditions are not fulfilled. In that case, the transfer term $\Pi_\ell$ is the only one contributing to the anomalous dissipation. Thus, sufficient conditions for the existence of a circulicity cascade are:
\begin{align} \label{eq:casc2}
\zeta_n^{v^d} > 1-\min( 2 \zeta_n^{j}, \zeta_n^{j^2})
\andd
\zeta_n^d >  1-\zeta_n^{vj}
\period
\end{align}
Other conditions may also lead to the fact that $\sub{\pc}$ vanishes in the limit $\mu \to 0$, $\ell \to 0$. For instance, $\bs{d}$ and $\bs{v}^d$  may simply be quantities belonging to the dissipative range that vanish when $\mu \to 0$. This is the case for $\bs{v}^d$ in the pseudo-compressible limit studied in \cite{sandoval1995,livescu2008,soulard2020}.

To conclude, the locality of the transfer term $\Pi_\ell$ obeys the same constraints as the locality of the energy transfer term studied in \cite{aluie2011}. We recall that locality requires the velocity and density exponents to satisfy:
\begin{align} \label{eq:local}
\zeta_n^v < n \andd \zeta_n^\rho >0.
\period
\end{align}
If equations \eqref{eq:casc}, \eqref{eq:casc2} and \eqref{eq:local} are verified then a local conservative cascade of circulicity takes place in the inertial range.

%=======================================================================
\section{Conclusion}
%=======================================================================
In \cite{eyink2018}, Eyink \& Drivas noted that ``it is not hard to find infinitely many anomalous balance relations in the ideal limit of turbulence, but most of them are not physically relevant and have no significant consequences''.
In this work, we started by raising a physical question: we asked whether a cascade of rotational motions  (whirls) exist in the inertial range of compressible turbulence.
This question was formulated in terms of a quantity called circulicity which carries information about  the large scale angular momentum of the flow.
We then used a point-splitting and a coarse-graining approach to analyse the properties of the circulicity.
The two approaches gave similar results but allowed to highlight different aspects.
Concerning the point-splitting approach, we derived a Monin-Yaglom equation that differs from previously derived expressions \cite{falkovich2010,galtier2011,wagner2012,banerjee2013,kritsuk2013,banerjee2014,banerjee2017,banerjee2018,andres2018,andres2019,ferrand2020,simon2021}.
In particular, it only involves quantities which can be reconstructed from the velocity and density fields, $\rho$ and $\bs{v}$. It does not require any knowledge about the energy or the pressure and is independent of the equation of state of the flow.
It is also valid whether the flow is fully compressible or is in a quasi-incompressible limit.
This Monin-Yaglom relation allowed to show that an ``effective'' cascade of circulicity exists when the flow is stirred with a solenoidal force, but not when the force is purely dilatational.
Besides, it also allowed to identify a weakly dilatational regime characterized by a Kolmogorov-Obukhov scaling and a scaling parameter given by a solenoidal dissipation.
In the opposite regime, it was shown how this Monin-Yaglom relation could be interpreted when strong shocks are present.
Finally, the practical relevance of this Monin-Yaglom was discussed. 

The coarse-graining point of view brought additional information and allowed to clarify several points. In particular, we were able to disentangle the role played by the actual transfer of circulicity and its subscale production.
This allowed to propose conditions under which a conservative and local circulicity cascade could take place.

%=======================================================================
% BIBLIOGRAPHY
%=======================================================================

\appendix
%=======================================================================
\section{Useful relationships for two-point correlations}
\label{sec:useful}
%=======================================================================
For some given quantities $A$, $B$ and $C$ with homogeneous statistics, we have:
\begin{align}
\rey{\Delta A \Delta B} =& 2 \rey{AB} - \rey{A B'} - \rey{A' B}
\coma
\\
\rey{\Delta A \Delta B \Delta C} =& \rey{A' (BC) } - \rey{A (BC)'}
+ \rey{B' (AC)} - \rey{B (AC)'} + \rey{C' (AB)} - \rey{C (AB)'}
\coma
\end{align}
\begin{align}
\rey{A'B} - \rey{AB'} =& 2 \rey{\Delta A \;\halfp{B}} = - 2 \rey{\Delta B\;\halfp{A}} =  \rey{\Delta A \;\halfp{B}} - \rey{\Delta B\;\halfp{A}}
\coma
\\
\rey{\partial_i'A' B} - \rey{\partial_i A B'} =& \partial_{r_i} (\rey{A' B} + \rey{A B'}) = - \partial_{r_i} \rey{\Delta A \Delta B}
\period
\end{align}
If $v_j$ satisfies $\partial_j v_j = \partial_j a_j$, we have:
\begin{align}
\partial_{r_j} (\rey{v_j' B' C} - \rey{v_j B C'}) =& - \partial_{r_j} \rey{\Delta v_j \Delta B \Delta C}
 - \partial_{r_j} (\rey{v'_j B C'} -  \rey{ v_j B' C} )
+ \partial_{r_j} (\rey{a'_j B C} -  \rey{ a_j B' C'} )
\coma
\\
\partial_{r_j}( \rey{v_j' B B'} - \rey{v_j B B'} ) =& - \frac{1}{2}\partial_{r_j} \rey{\Delta v_j \Delta B^2} + \frac{1}{2} \partial_{r_j} (\rey{a'_j B^2} -  \rey{ a_j B'^2} )
\end{align}
We also recall that homogeneity implies that:
\begin{align}
\rey{\partial'_j( \cdot)} = - \rey{\partial_j(\cdot)} = \partial_{r_j} ( \rey{\phantom{(} \cdot\phantom{)} })
\period
\end{align}

%=======================================================================
\section{Evolutions of the angular momentum and of the   dilatational part of the linear momentum}
%=======================================================================

The evolution of the angular momentum $\bs{\Omega}=\epsilon_{ijk} \partial_j (\rho v_k)$ is deduced from Eq. \eqref{eq:ns_rhou}, knowing that $\partial_j(\rho _j v_k) = v_j \partial_k(\rho v_j) + \rho v_k \partial_jv_j - \epsilon_{kpq} v_p \Omega_q$:
\begin{align}
  \partial_t \Omega_i + v_j \partial_j \Omega_i = \Omega_j \partial_j v_i -
\Omega_i \partial_j v_j 
  + \epsilon_{ijk} \partial_k\left( \frac{\rho v_lv_l}{2} \right) \frac{\partial_j \rho}{\rho}
- \epsilon_{ijk} \partial_j\Big( \rho v_k \partial_lv_l  \Big)
+ \mu \partial^2_{jj} \omega_i
+ \epsilon_{ijk} \partial_j f_k^s
  \period
\end{align}
This equation can be compared against the vorticity equation:
\begin{align}
  \partial_t \omega_i + v_j \partial_j \omega_i = \omega_j \partial_j v_i -
\omega_i \partial_j v_j 
  + \epsilon_{ijk} \frac{\partial_k p}{\rho} \frac{\partial_j \rho}{\rho}
+ \frac{\mu}{\rho} \partial^2_{jj} \omega_i + \epsilon_{ijk} \frac{\partial_l \sigma_{kl}}{\rho} \frac{\partial_j \rho}{\rho}
+ \epsilon_{ijk} \partial_j (f_k/\rho)
\period
\end{align}
We note that similar stirring terms are present in both equations, while the standard baroclinic torque acting in the vorticity equation is replaced by a torque involving the dynamic pressure $\rho |\bs{v}|^2/2$ in the angular momentum equation.

As for $\bs{d}$, its evolution is given by:
\begin{align}
& \partial_t d_i  = -\partial_i \pi^d - \partial_j \sigma_{ij}^d + f_i^d
\coma
\\
\nonumber
\with &
\pi^d = p- \pis
\coma
\sigma_{ij}^d = - (\eta + \frac{4}{3}\mu) \partial_k v_k \delta_{ij}
\coma
f_i^d = - \partial_i \phi^f 
\coma
\end{align}
and $\partial_{jj}\phi^f = - \partial_j f_j$.
This equation is readily integrated and shows that the scalar potential $\phi$, such that $d_i = -\partial_i \phi$, is directly related to the dilatational pressure $\pi^d$ in addition to forcing and viscous terms:
\begin{align}
\phi(\x,t) = \phi_0(t) + \int_0^t  \pi^d(\x,t') \ud t'  +  \int_0^t \phi^f(\x,t')  - (\eta + \frac{4}{3}\mu) \partial_k v_k(\x,t') \ud t'
\coma
\end{align}
with $\phi_0$ an arbitrary function of time.

%=======================================================================
\section{Karman-Howarth equation for the momentum}
\label{app:momentum}
%=======================================================================
The Karman-Howarth equation for the linear momentum $\jjv = \rho \bs{v}$ can be found in \cite{eyink2018}. Its expression, derived from Eq. \eqref{eq:ns_rhou}, is recalled here:
\begin{subequations}
\begin{align}
\label{eq:dqdq}
& \partial_t \rey{\Delta \jj_i \Delta \jj_i}(\R,t) = - \partial_{r_j} \mathscr{F}^\jj_j(\R,t) - 4 \Einj^\jj(t)  +  \partial^2_{r_jr_k}\mathcal{D}_{jk}^{\jj}(\R,t)  + 2 \rey{\Delta \jj_i \Delta f_i}
\coma
\end{align}
with:
\begin{align}
& \mathscr{F}^\jj_j(\R,t)= \rey{\Delta v_j \Delta \jj_i \Delta \jj_i} 
+ 2  \partial_{r_j} \Big( \rey{\Delta \Phi \Delta p}  - \frac{1}{2} \rey{\Delta \varphi^v \Delta (\jj_i\jj_i) } \Big)
\coma
\\
& \Einj^\jj(t) = - \left( \rey{\sigma_{ij} \partial_j \jj_i} +  \rey{p \partial_j \jj_j} - \frac{1}{2} \rey{\jj_i\jj_i \partial_jv_j} 
\right)
\coma
\\
& \mathcal{D}_{jk}^\jj(\R,t) = 2 \mu \rey{\Delta v_i \Delta \jj_i} \delta_{jk}+ 2 (\eta+\frac{\mu}{3}) \rey{\Delta v_j \Delta \jj_k}
\period
\end{align}
\end{subequations}

\end{document}